\documentclass[twocolumn,showkeys,showpacs,preprintnumbers,prd,superscriptaddress,nofootinbib,aps,10pt,floatfix]{revtex4-1}
\usepackage{graphicx,epsf,bm,amsmath,amsfonts,amssymb,epstopdf,natbib,color,verbatim,multirow,bm,mathtools,mathrsfs,braket,bbold,xcolor}
\usepackage{hyperref}
\usepackage[normalem]{ulem}
\hypersetup{colorlinks=true,urlcolor=blue,citecolor=blue,linkcolor=blue,menucolor=blue,anchorcolor=blue,filecolor=blue}
\date{\today}

\newcommand{\beff}{\beta_{\mathrm{eff}}}
\newcommand{\csTwo}{\mathrm{cS2}}

\begin{document}

\title{Probing Solar Chameleons with XENONnT Ionization-Only Data}

\author{Guan-Wen Yuan}
\email{guanwen.yuan@unitn.it; \\ guanwen.yuan@ustc.edu.cn.}
\affiliation{Department of Physics, University of Trento, Via Sommarive 14, 38123 Povo (TN), Italy \looseness=-1}
\affiliation{Trento Institute for Fundamental Physics and Applications (TIFPA)-INFN, Via Sommarive 14, 38123 Povo (TN), Italy \looseness=-1}
\affiliation{Department of Astronomy, University of Science and Technology of China, Hefei, Anhui 230026, China}
\affiliation{School of Astronomy and Space Science, University of Science and Technology of China, Hefei, Anhui 230026, China}

\begin{abstract}
Screened scalar fields provide a possible connection between dark energy and laboratory-scale new physics while remaining compatible with existing fifth-force constraints. Solar production offers a direct probe of these models, and ionization-only measurements extend the sensitivity of liquid-xenon detectors into the sub-keV regime. 
We present the first search for solar chameleons using low-energy S2-only data, based on the 7.83 tonne-year XENONnT ionization-only data set. The predicted electronic-recoil spectrum is folded through the run-dependent detector response and tested directly in corrected-S2 ($\csTwo$) space. A joint binned likelihood combines the seven $\csTwo$ bins of science runs SR0, SR1, and SR2. We sample the five-dimensional parameter space $\{\beta_e,\beta_\gamma,M_e,\Lambda,n\}$ and derive posterior constraints on the conformal and disformal electron couplings. The data extend the recoil-energy reach to approximately $0.04~{\rm keV}_{ee}$, where the few-electron detector response and instrumental backgrounds must be modeled explicitly. In the disformal-dominated regime, the result is expressed in terms of $\beta_{\rm eff}\equiv\beta_\gamma({\rm eV}/M_e)^4$. We obtain the upper limit $\log_{10}\beta_{\rm eff}<-3.64$. Although the present analysis reaches a lower energy threshold than our previous XENONnT electronic-recoil analysis, the resulting constraint is weaker, primarily because the disformal absorption rate decreases rapidly toward low recoil energies, where the background rate also becomes larger. 
This analysis establishes low-energy ionization data as a new experimental channel for testing solar chameleons and screened dark-energy models.
\end{abstract}

\maketitle

\section{Introduction}
\label{sec:introduction}

The nature of dark energy remains one of the central open problems in modern cosmology~\cite{Riess:1998cb,Perlmutter:1998np, Nojiri:2006ri, Frieman:2008sn, Bamba:2012cp, Huterer:2017buf}. A broad class of dynamical dark-energy scenarios introduces an additional light scalar field whose cosmological evolution can drive or contribute to the observed accelerated expansion of the Universe~\cite{Ratra:1987rm,Wetterich:1987fm,Caldwell:1997ii,Zlatev:1998tr,Copeland:2006wr,Tsujikawa:2013fta,Odintsov:2020nwm,Giare:2024sdl,Kaneta:2025kcn,Lin:2025gne, Cai:2025mas, Giare:2026oti,Gomez-Valent:2026ept,Huang:2025som,Gonzalez-Fuentes:2026rgu}. 
Such light scalars are, however, strongly constrained by laboratory, solar system, and astrophysical tests of fifth forces ~\cite{Riotto:2000kh,Torres:2000dw,Hu:2000ke,Amendola:2005ad,Adelberger:2003zx,Kapner:2006si, Serebrov:2009zz, Chiow:2011zz, Brito:2015oca,Hees:2016gop, Safronova:2017xyt,SimonsObservatory:2018koc, Schive:2014hza,Kouvaris:2019nzd,Chen:2021lvo, SimonsObservatory:2019qwx,Yuan:2020xui, Roy:2021uye,Odintsov:2022cbm,Chen:2022nbb,Poulin:2023lkg,Kading:2023mdk,Alesini:2023qed,Kading:2024jqe,Calza:2025yfm, Lyu:2025nsd}. Screening mechanisms provide a possible way to reconcile a light cosmological scalar with these constraints by suppressing its observable effects in dense environments while retaining nontrivial dynamics on cosmological scales.

Among the best-studied screening scenarios~\cite{Khoury:2010xi,Brax:2013ida,Joyce:2014kja, Sakstein:2018fwz, Baker:2019gxo, Feleppa:2025vop} is the chameleon mechanism, in which the effective mass of the scalar field depends on the ambient matter density(see, e.g.\ Refs.~\cite{Brax:2004ym,Brax:2004qh,Capozziello:2007eu,Brax:2008hh,Brax:2010kv,Gannouji:2010fc,Wang:2012kj,Erickcek:2013dea,Elder:2016yxm,Brax:2016did,Burrage:2017shh,Burrage:2018pyg,Katsuragawa:2019uto,Sakstein:2019qgn,Desmond:2019ygn,Hartley:2019wzu,Karwal:2021vpk,Katsuragawa:2021wmw,Dima:2021pwx,Benisty:2021cmq,Tamosiunas:2021kth,Briddon:2021etm,Brax:2021owd,Yuan:2022cpw,Chakrabarti:2022zvv,Tamosiunas:2022tic,Brax:2022olf,Benisty:2022lox,Boumechta:2023qhd,Elder:2023oar,Benisty:2023dkn,Paliathanasis:2023ttu,Paliathanasis:2023dfz,Benisty:2023vbz,Briddon:2023ayq,Hogas:2023pjz,Zaregonbadi:2023vcv,Benisty:2023clf,Fischer:2023eww,Kumar:2024ylj,Fischer:2024coj,Baez-Camargo:2024jia,Fischer:2024eic,Paliathanasis:2024sle,Fischer:2024gni,Pizzuti:2024hym,Nojiri:2025low,Zaregonbadi:2025ils,Feleppa:2025clx,Neckam:2025kip,Feleppa:2025vop}). The scalar can therefore become heavy and
short-ranged in terrestrial environments, while remaining light in
sufficiently dilute regions~\cite{Brax:2010xq, Brax:2011wp,Cai:2021wgv}. Generalized chameleon models can couple
not only conformally to matter and photons, but also through
derivative or disformal operators. These interactions open a variety
of laboratory~\cite{Burrage:2016bwy, Homma:2019rqb, Yuan:2022nmu, Bashyam:2026aiz} and astrophysical probes~\cite{Chang:2010xh,Iorio:2011ay,Llinares:2012ds,Llinares:2013jua,Gronke:2014gaa,Gronke:2015ama,Santos:2016rdg,Katsuragawa:2016yir,Katsuragawa:2017wge,Olmo:2019flu,KumarPoddar:2020kdz,Straight:2020zke,Tsai:2021irw,Banks:2026bso} that are complementary to
traditional fifth-force experiments.  

The Sun provides a particularly interesting environment for testing
such models, because the chameleons can be produced in the hot solar plasma
through their coupling to photons and subsequently reach terrestrial detectors. 
Previous studies have discussed the production from photon--chameleon conversion in solar magnetic fields and the Primakoff-like mechanism in the screened electric fields of electrons and ions, these channls can can make an important contribution to the solar chameleon flux~\cite{Vagnozzi:2021quy, OShea:2024jjw,Yuan:2025twx}. 
Once produced, these solar chameleons can interact with electrons in low-background terrestrial detectors. Liquid-xenon time-projection
chambers are particularly well suited to such searches because they
combine large target masses with sensitivity to low-energy electronic
recoils. In Ref.~\cite{Yuan:2025twx}, we performed a dedicated search
for solar chameleons using XENONnT electronic recoil data. That study
included both magnetic and Primakoff solar production and considered
conformal and disformal chameleon--electron interactions. We found
that the resulting event rate is controlled by the combination
of the photon coupling and the electron disformal scale
\begin{equation}
\beff \equiv \beta_\gamma
\left( \frac{{\rm eV}}{M_e} \right)^4 ,
\end{equation}
which captures the approximate degeneracy between solar production
and terrestrial detection.

Recently, the XENON Collaboration has released a dedicated ionization-only
(S2-only) analysis of XENONnT data with a total exposure of
$7.83~{\rm tonne\,yr}$ accumulated over three science runs~\cite{XENON:2026gxb}.
Unlike conventional S1--S2 searches, the ionization-only strategy
remains sensitive when the prompt scintillation signal becomes too
small to be efficiently detected, extending the electronic-recoil (ER)
sensitivity down to approximately $ 0.04~{\rm keV}_{ee} \lesssim E_R \lesssim 0.7~{\rm keV}_{ee}$.
The collaboration reports no significant excess over the expected
background and provides public data and detector-response information
for reinterpretation~\cite{XENON:2026gxb}. The S2-only
search combines SR0, SR1, and SR2 and performs the statistical
inference directly in the corrected-S2 ($\csTwo$) observable rather
than in reconstructed recoil-energy space. This low-threshold
approach is particularly relevant for models such as solar
chameleons, whose predicted ER spectrum is strongly weighted toward sub-keV energies, a regime that has not previously been explored.

The reinterpretation of the ionization-only data requires a
substantially different treatment from that employed in our previous
analysis. The theory calculation predicts a differential recoil
spectrum in the true deposited ER energy,
$dR_0/dE_R$, whereas the experimental analysis is performed in
$\csTwo$ space. At these low energies, only a few ionization electrons
are produced, and fluctuations in the charge yield, electron
extraction, and S2 generation lead to a broad, run-dependent mapping
between $E_R$ and $\csTwo$. Consequently, the signal prediction must
be propagated through the XENONnT ER response matrices
before it can be compared with the data. In addition, the three
science runs have different exposures, thresholds, and instrumental
background compositions, and therefore need to be treated separately
in the likelihood. The collaboration itself performs the S2-only
inference in the cS2 dimension and treats SR0, SR1, and SR2
separately before combining them statistically. 

In this work, we revisit the direct detection of solar chameleons
using the public XENONnT ionization-only data. We retain the solar
production and chameleon--electron interaction framework developed in
Ref.~\cite{Yuan:2025twx}, but replace the reconstructed-energy
treatment by a response-level analysis in the experimentally measured
$\csTwo$ observable. The solar chameleon recoil spectrum is folded
with the run-dependent XENONnT ER-to-$\csTwo$ response matrices to
construct signal templates for the seven analysis bins of each of
SR0, SR1, and SR2. These are then combined with the released
background templates in a joint likelihood.
We explore the full five-dimensional chameleon parameter space
$\left\{ \beta_e,\, \beta_\gamma,\, M_e,\, \Lambda,\,  n \right\}$,
using Bayesian inference. We present the corresponding marginalized
posterior distributions, identify a representative
maximum-a-posteriori signal realization, and derive the updated
constraint in the $\beff$--$\beta_e$ plane after marginalizing over
the remaining chameleon and nuisance parameters. This analysis
therefore tests solar chameleons in a recoil energy regime that was
largely inaccessible to the higher-threshold electronic recoil spectrum used in our
previous study.

The remainder of this paper is organized as follows.
In Sec.~\ref{sec:solarchameleons}, we briefly summarize the generalized
chameleon framework, solar production mechanisms, and the resulting
chameleon-electronic recoil spectrum. In
Sec.~\ref{sec:xenon_analysis}, we describe the XENONnT ionization-only
data, the ER-to-$\csTwo$ detector response, the background model, and
the statistical analysis. Section~\ref{sec:results} presents the
posterior constraints and the resulting limits on the effective
chameleon couplings. We discuss the implications of the new
low-threshold constraints and summarize our conclusions in
Sec.~\ref{sec:discussion}.

\section{Solar Chameleon Signal in XENON${\bf nT}$}
\label{sec:solarchameleons}

In this section, we summarize the ingredients required to predict the electronic-recoil (ER) signal induced by solar chameleons in XENONnT.
A detailed discussion of the generalized chameleon framework, the solar production calculation, and the corresponding detection channels was presented in Ref.~\cite{Yuan:2025twx}. Therefore, we restrict the discussion here to the aspects directly relevant to the present S2-only analysis.

\subsection{Chameleon framework and relevant couplings}
\label{sec:framework}

We consider a generalized chameleon scalar field $\phi$ with the
inverse-power-law self-interaction potential
\begin{equation}
V_{\rm self}(\phi) =\Lambda^4
\left[ 1+\left(\frac{\Lambda}{\phi}\right)^n \right],
\label{eq:potential}
\end{equation}
where $\Lambda$ sets the characteristic energy scale of the potential
and $n>0$ is the power-law index. Values close to the dark-energy scale,
$\Lambda_{\rm DE}\simeq 2.4~{\rm meV}$, are of particular interest,
although in the present analysis we allow $\Lambda$ to vary.

For the phenomenology considered here, the relevant interactions are
the conformal coupling to photons, together with the conformal and
disformal couplings to electrons. Schematically, these interactions can
be written as
\begin{equation}
\mathcal{L}_{\rm int}
\supset
-\frac{\beta_\gamma}{4M_{\rm Pl}}\,
\phi F_{\mu\nu}F^{\mu\nu}
-\frac{\beta_e m_e}{M_{\rm Pl}}\,
\phi\bar e e
+\frac{1}{M_e^4}
\partial_\mu\phi\,\partial_\nu\phi\,T_e^{\mu\nu},
\label{eq:interaction}
\end{equation}
where $M_{\rm Pl}$ is the reduced Planck mass, $\beta_\gamma$ is the
dimensionless conformal coupling to photons, $\beta_e$ is the
conformal coupling to electrons, and $M_e$ denotes the electron
disformal scale. Together with $\Lambda$ and $n$, these quantities
define the five-dimensional parameter space explored below,
\begin{equation}
\boldsymbol{\Theta} =
\left\{ \beta_e,\, \beta_\gamma,\, M_e,\, \Lambda,\, n \right\}.
\label{eq:model_parameters}
\end{equation}

The defining feature of the chameleon mechanism is that the scalar
acquires an environment-dependent effective mass. In a medium of
matter density $\rho$, and neglecting the electromagnetic contribution
to the effective potential, the mass at the density-dependent minimum
is approximately
\begin{equation}
m_{\rm eff}^2(\rho)\simeq n(n+1)\Lambda^{4+n}
\left(\frac{\beta_e\rho} {nM_{\rm Pl}\Lambda^{4+n}}
\right)^{\frac{n+2}{n+1}},
\label{eq:mass_eff}
\end{equation}
where $\beta_e$ denotes the density-weighted conformal matter coupling.
In the restricted parameterization used here, only the electron
conformal coupling is retained, and we identify the density-weighted
coupling entering Eq.~\eqref{eq:mass_eff} with $\beta_e$. 
The increase of $m_{\rm eff}$ with ambient density is responsible for
the screening of the scalar-mediated fifth force. For the solar
chameleons of interest here, the same density dependence determines
whether a mode can propagate through the solar plasma and through the
materials encountered before reaching the active xenon target. We
apply the same propagation requirements as in
Ref.~\cite{Yuan:2025twx}; within the parameter region relevant to the
present analysis, these conditions do not introduce an additional
observable parameter beyond those in Eq.~\eqref{eq:model_parameters}.

\subsection{Solar chameleon production}
\label{sec:solar_production}

The hot solar plasma provides an efficient source of relativistic
chameleons through the photon coupling in
Eq.~\eqref{eq:interaction}. We include the two transverse-photon
production mechanisms:
Primakoff production in the screened electric fields of electrons and
ions, and photon--chameleon conversion in the solar magnetic field.
The total differential flux at Earth is therefore
\begin{equation}
\frac{d\Phi_\phi}{d\omega} = \frac{1}{4\pi D_\odot^2}
\left(\frac{d\dot N_{\rm P}}{d\omega}+\frac{d\dot N_{B}}{d\omega} \right),
\label{eq:flux}
\end{equation}
where $\omega$ is the chameleon energy and $D_\odot$ is the
Sun--Earth distance.

For Primakoff production,
$\gamma+Ze\rightarrow Ze+\phi$, the screened Coulomb fields of the
charged particles in the solar plasma provide the required momentum
transfer. The production spectrum can be written as
\begin{align}
\frac{d\dot N_{\rm P}}{d\omega} 
={}& \frac{\beta_\gamma^2\alpha} {8\pi M_{\rm Pl}^2}
\int_0^{R_\odot} dr\, r^2 \frac{1}{e^{\omega/T(r)}-1} \nonumber\\
&\times \frac{\omega^2 k_\phi}{k_\gamma} \,\mathcal{I}(u,v) \sum_i Z_i^2 n_i(r),
\label{eq:primakoff_rate}
\end{align}
where $k_\phi$ and $k_\gamma$ are the chameleon and in-medium photon
momenta, respectively, $n_i$ and $Z_i$ are the number density and
charge of species $i$, and $\mathcal{I}(u,v)$ contains the angular
integral and Debye-screening dependence. We evaluate these quantities
using the AGSS09~\cite{Asplund:2009fu} solar temperature, density, and elemental-abundance
profiles, following the numerical treatment developed in
Refs.~\cite{OShea:2024jjw,Yuan:2025twx}.

The second contribution arises from photon--chameleon conversion in
the bulk solar magnetic field. Its rate depends on the radial
magnetic-field profile, the plasma-induced photon mass, and the photon
absorption width. 
The resulting spectrum is given by the following expression:
\begin{align}
\frac{{\rm d}\dot{N}_B}{{\rm d}\omega} = &\nonumber 
\frac{2\beta_{\gamma}^2}{\pi M_{\text{Pl}}^2}\int_0^{R_{\odot}}{\rm d}r\,r^2B_{\perp}^2(r) \\
&\times \frac{\omega(\omega^2-m^2)^{3/2}}{(m_{\gamma}^2 - m^2)^2+(\omega\Gamma_{\gamma})^2}\frac{\Gamma_{\gamma}}{e^{\omega/T}-1}\,,
\label{eq:productionmagnetic}
\end{align}
where $B_{\perp}$ denotes the magnetic field component perpendicular to the propagation direction, and $m_{\gamma}$ is the plasma-induced effective photon mass, whereas $\Gamma_{\gamma}(\omega,r)$ accounts for a number of photon production and absorption processes within the solar medium. 
We employ the same radiative-zone, tachocline, and
convective-zone magnetic-field model adopted in
Ref.~\cite{Yuan:2025twx}. Both the Primakoff and magnetic
conversion amplitudes are proportional to $\beta_\gamma$, such that
their production rates satisfy
\begin{equation}
\frac{d\dot N_{\rm P}}{d\omega}, \frac{d\dot N_B}{d\omega}
\propto \beta_\gamma^2.
\label{eq:production_scaling}
\end{equation}
We factor out this common normalization when displaying the solar
spectrum.

\begin{figure}[!t]
\centering
\includegraphics[width=0.95\linewidth]{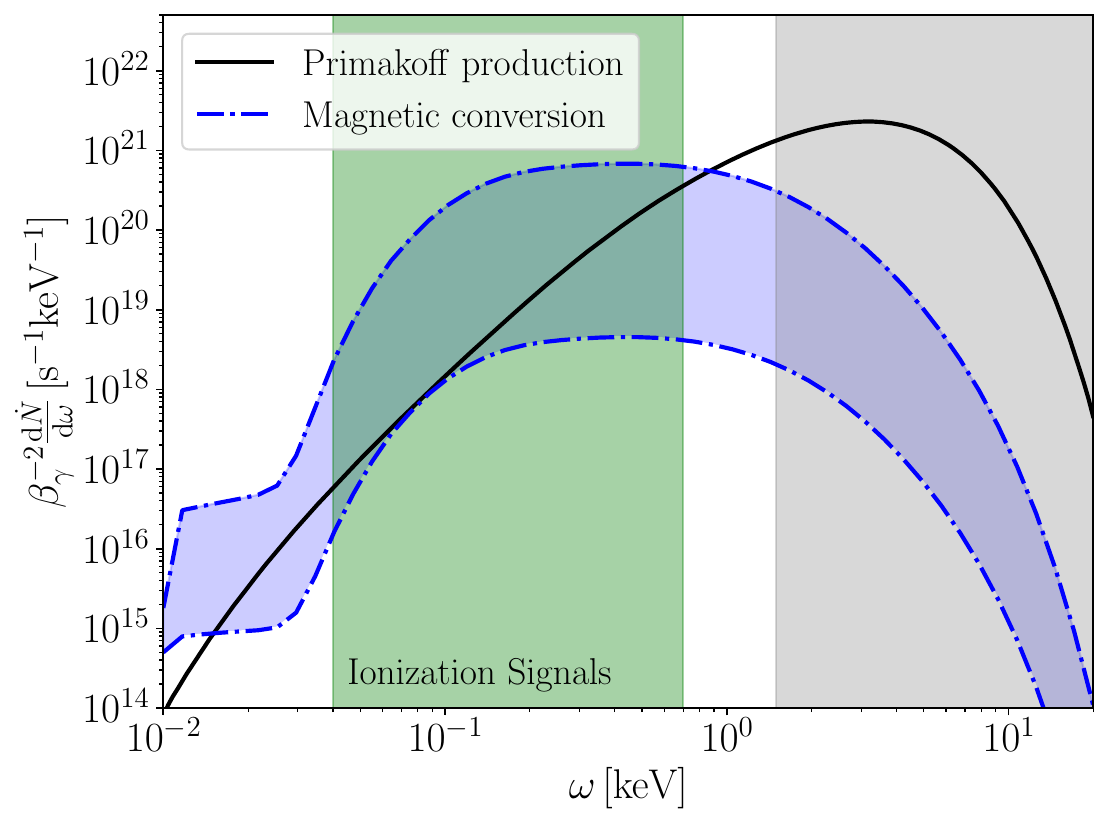}
\caption{Solar chameleon production spectrum normalized by $\beta_\gamma^2$. The solid black curve shows Primakoff production in the screened electric fields of electrons and ions; the blue band shows photon--chameleon conversion in the solar magnetic field and spans the assumed field-strength uncertainty. We use the AGSS09 solar model and the production calculation with $\beta_e=10^2$, $\Lambda=2.4~{\rm meV}$, and $n=1$. The green band marks the electronic-recoil interval used in the XENONnT ionization-only analysis.}
\label{fig:spectrum}
\end{figure}

Fig.~\ref{fig:spectrum} shows the production spectrum after dividing
out the overall $\beta_\gamma^2$ dependence. The Primakoff contribution dominates over
magnetic conversion throughout most of the energy range relevant for
direct detection, while the magnetic component increases in relative
importance toward lower energies. The Primakoff contribution is less
sensitive to uncertainties in the solar
magnetic-field model, since it depends primarily on the well-measured
thermodynamic and composition profiles of the Sun.

The production model includes transverse plasma excitations only.
Longitudinal plasma modes may
provide an additional low-energy contribution to the solar chameleon
flux, but a dedicated calculation for the generalized chameleon model
considered here is not presently available. This contribution was
negligible for the ${\cal O}({\rm keV})$ analysis of
Ref.~\cite{Yuan:2025twx}. Since the present S2-only search extends to
sub-keV ER energies, however, its possible impact deserves further study.
The limits below therefore apply to the transverse-production model.
An additional positive production contribution would strengthen them.

\subsection{Chameleon--electron recoil spectrum}
\label{sec:recoil_spectrum}

After escaping the Sun, relativistic chameleons can reach the Earth and
deposit their energy in the liquid-xenon target through interactions
with atomic electrons. For the couplings in
Eq.~\eqref{eq:interaction}, the interaction cross section per unit
target mass is the incoherent sum of a disformal contribution and a
conformal, photoabsorption-like contribution~\cite{OShea:2024jjw},
\begin{align}
\sigma_{\phi e}(E_R) \equiv{}& \sigma_{\phi e}^{\rm dis}(E_R) +\sigma_{\phi e}^{\rm conf}(E_R) \\
={}& N_{\rm Xe}\frac{m_e^2 E_R^4} {8\pi^2 M_e^8} +\frac{\beta_e^2 E_R^2}
{2\pi\alpha M_{\rm Pl}^2} \sigma_{\rm photo}(E_R)  \nonumber,
\label{eq:cross-section}
\end{align}
where $N_{\rm Xe}$ is the number of xenon atoms per unit detector mass,
$m_e$ is the electron mass, $\alpha$ is the fine-structure constant,
and $\sigma_{\rm photo}$ is the xenon photoelectric cross section. The
two terms originate from distinct operators and do not interfere.

Since the chameleon mass inside the XENONnT target is negligible
relative to the recoil energies of interest, the absorbed chameleon
energy can be identified with the deposited ER energy,
$\omega\simeq E_R$. The differential event rate before detector
response effects is therefore
\begin{equation}
\frac{dR_0}{dE_R} =\sigma_{\phi e}(E_R) \frac{d\Phi_\phi}{dE_R}.
\label{eq:true-rate}
\end{equation}
Eq.~\eqref{eq:true-rate} defines the theory-level spectrum that
will be propagated into the experimentally measured corrected-S2
($\csTwo$) observable in Sec.~\ref{sec:xenon_analysis}.

The different parameter dependences of production and detection lead
to a useful approximate degeneracy. Since the solar flux scales as
$\beta_\gamma^2$, whereas the dominant disformal detection cross
section scales as $M_e^{-8}$, the corresponding event rate obeys
\begin{equation}
\frac{dR_0^{\rm dis}}{dE_R} 
\propto \beta_\gamma^2 M_e^{-8}.
\label{eq:rate_scaling}
\end{equation}
Hence, we introduce the dimensionless effective coupling $\beff \equiv \beta_\gamma \left(\frac{\mathrm{eV}}{M_e}\right)^4$,
in terms of which the disformal contribution scales as
\begin{equation}
\frac{dR_0^{\rm dis}}{dE_R} \propto \beff^2.
\end{equation}
This combination captures the dominant degeneracy between solar
production and terrestrial detection and is used below to present the
final constraints. Throughout the
full inference we retain the original five chameleon parameters in
Eq.~\eqref{eq:model_parameters}, allowing the conformal contribution
controlled by $\beta_e$, as well as the dependence on $\Lambda$ and
$n$, to be consistently marginalized over.

\section{XENON${\bf nT}$ Ionization-Only Analysis}
\label{sec:xenon_analysis}

We now describe the procedure used to confront the solar-chameleon
electronic-recoil spectrum derived in Sec.~\ref{sec:solarchameleons}
with the recently released XENONnT ionization-only data.
The essential difference from our previous analysis
in Ref.~\cite{OShea:2024jjw,Yuan:2025twx} is that the theoretical spectrum is no
longer compared directly with a reconstructed recoil-energy spectrum.
Instead, it is propagated through the detector response into the
experimentally measured corrected-S2 observable, $\csTwo$, and the
statistical analysis is performed in this observable space.

\subsection{Data set and observable}
\label{sec:dataset}

The XENONnT ionization-only search combines three science runs,
denoted SR0, SR1, and SR2, collected between May 2021 and March 2025.
The corresponding live times are 110.2, 177.8, and 291.5 days,
respectively, yielding exposures of

\begin{equation}
\begin{aligned}
\mathcal{E}_{\rm SR0}&=1.50~{\rm tonne\,yr}, \\
\mathcal{E}_{\rm SR1}&=2.44~{\rm tonne\,yr}, \\
\mathcal{E}_{\rm SR2}&=3.89~{\rm tonne\,yr}, 
\end{aligned}
\end{equation}
and a total exposure of $\mathcal{E}_{\rm tot}=7.83~{\rm tonne\,yr}$.
The principal characteristics of the three data-taking periods are summarized in Table~\ref{tab:runs}.

\begin{table}[t]
\caption{
Live times and exposures of the three XENONnT science runs used in
the ionization-only analysis~\cite{XENON:2026gxb}.
}
\label{tab:runs}
\begin{ruledtabular}
\begin{tabular}{lcc}
Run & Live time [day] & Exposure [tonne yr] \\
\hline
SR0 & 110.2 & 1.50 \\
SR1 & 177.8 & 2.44 \\
SR2 & 291.5 & 3.89 \\
\hline
Total & 579.5 & 7.83
\end{tabular}
\end{ruledtabular}
\end{table}

In a dual-phase xenon time-projection chamber, an electronic recoil
produces both prompt scintillation photons (S1) and ionization
electrons. At sufficiently low recoil energies the S1 signal frequently
falls below threshold, whereas even a small number of extracted
electrons can still generate a detectable proportional-scintillation
signal (S2). The S2-only strategy therefore extends the sensitivity of
XENONnT substantially below that of conventional S1--S2 analyses.

Candidate events are selected with an S2 area between approximately
100 and 500 photoelectrons (PE) in SR0 and SR1, and between
120 and 500 PE in SR2, corresponding to approximately $3$--$16$
and $4$--$16$ extracted electrons, respectively. The higher threshold
in SR2 reflects changes in the low-level event reconstruction designed
to suppress single-electron pileup. The inference variable is the
cS2 area,
\begin{equation}
80~{\rm PE} \leq \csTwo \leq 500~{\rm PE},
\end{equation}
where corrections for spatial and temporal variations of the
single-electron gain and extraction efficiency have been applied.
This ionization-only analysis is sensitive to electronic-recoil
energies approximately in the interval $0.04~{\rm keV}_{ee} \lesssim E_R
\lesssim 0.7~{\rm keV}_{ee}$.

\begin{table}[tb]
\caption{The seven $\csTwo$ bins and observed event counts in each science run.}
\label{tab:bins}
\begin{ruledtabular}
\begin{tabular}{ccccc}
Bin $k$ & $\csTwo$ (PE) & $n_{{\rm SR0},k}$ & $n_{{\rm SR1},k}$ & $n_{{\rm SR2},k}$ \\
\hline
1 & [80, 140] & 220 & 239 & 239 \\
2 & [140, 200] & 171 & 243 & 391 \\
3 & [200, 260] & 85 & 160 & 222 \\
4 & [260, 320] & 48 & 90 & 107 \\
5 & [320, 380] & 28 & 69 & 83 \\
6 & [380, 440] & 21 & 43 & 37 \\
7 & [440, 500] & 10 & 20 & 28 \\
\hline
Total & [80, 500] & 583 & 864 & 1107 \\
\end{tabular}
\end{ruledtabular}
\end{table}

For our reinterpretation, we use the publicly released XENONnT
science data and background templates in seven $\csTwo$ bins for
each of the three science runs. The statistical data vector is
therefore composed of 21 independent run--bin combinations,
\begin{equation}
\mathcal{D}= \left\{N_{ri}^{\rm obs}\right\},
\quad r\in\{{\rm SR0,SR1,SR2}\},
\quad i=1,\ldots,7 ,
\label{eq:data_vector}
\end{equation}
where $N_{ri}^{\rm obs}$ denotes the observed number of events in
the $i$th $\csTwo$ bin of science run $r$.

\subsection{Detector response from electronic recoil to \texorpdfstring{$\csTwo$}{cS2}}
\label{sec:response}

The theory calculation in Sec.~\ref{sec:recoil_spectrum} predicts the
differential event rate as a function of the true deposited
electronic-recoil energy, ${dR_0}/{dE_R}$.
This quantity cannot be compared directly with the S2-only data.
At sub-keV energies the number of ionization electrons is small and
subject to sizeable statistical fluctuations, and the mapping between
$E_R$ and the observed S2 signal depends on the charge yield, electron
survival and extraction, single-electron gain, detector resolution,
event reconstruction, and analysis selections.

\begin{figure*}[htbp]
\centering
\includegraphics[width=0.98\linewidth]{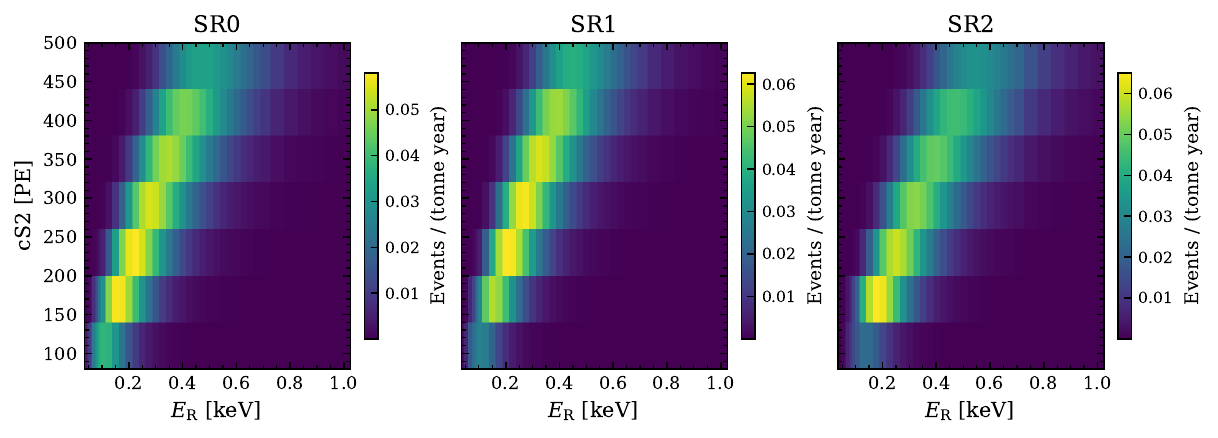}
\caption{
Electronic recoil response matrices used to map the true recoil
energy $E_R$ onto the corrected-S2 observable for SR0 (left),
SR1 (middle), and SR2 (right). Each vertical slice describes the
distribution in $\csTwo$ generated by a monoenergetic electronic
recoil. The matrices include the selection acceptance and are
normalized per unit exposure. Common axes and color scales expose the
run-dependent thresholds and resolution. Eq.~\eqref{eq:response_folding}
integrates these responses over the seven bins listed in
Table~\ref{tab:bins}.}
\label{fig:response_matrix}
\end{figure*}

We account for these effects using the run-dependent electronic-recoil
response matrices of XENONnT. We denote $\mathcal{R}_{ri}(E_R)$ as the response of science run $r$ for an electronic recoil of true energy $E_R$ to be reconstructed in the $i$th $\csTwo$ bin.
The expected solar-chameleon signal in that bin is then
\begin{equation}
S_{ri}(\boldsymbol{\Theta})=\mathcal{E}_{r} \int dE_R\,
\mathcal{R}_{ri}(E_R) \frac{dR_0(E_R|\boldsymbol{\Theta})}{dE_R},
\label{eq:response_folding}
\end{equation}
where $\boldsymbol{\Theta} = \{\beta_e,\beta_\gamma,M_e,\Lambda,n\}$
denotes the chameleon parameters.
For the numerical calculation, Eq.~\eqref{eq:response_folding} is
evaluated on the energy grid of the released response matrices. In
discretized form,
\begin{equation}
S_{ri} \simeq \mathcal{E}_r \sum_j \mathcal{R}_{rij}
\left. \frac{dR_0}{dE_R} \right|_{E_j} \Delta E_j ,
\label{eq:response_discrete}
\end{equation}
where $j$ labels the true-ER-energy bins and
$\mathcal{R}_{rij}$ is the corresponding response-matrix element.
We interpolate the theory spectrum onto the same energy grid before
performing the convolution.

The three response matrices are shown in
Fig.~\ref{fig:response_matrix}. As expected, low-energy recoils
populate the few-electron region close to threshold, while increasing
$E_R$ shifts the reconstructed distribution toward larger $\csTwo$.
The finite width of the response at fixed recoil energy illustrates
why a deterministic conversion between $E_R$ and $\csTwo$ is
insufficient in this regime.

The response matrices provide the complete mapping between the theory-level recoil
spectrum and the selected S2-only signal sample. We therefore do not
apply the energy-dependent efficiency function used in our previous
reconstructed-energy analysis as an additional multiplicative factor.
Doing so would count part of the detector acceptance twice. This
replacement,
\begin{equation}
\epsilon(E_R)\, G(E_{\rm obs},E_R)
\quad\longrightarrow\quad
\mathcal{R}_{ri}(E_R),
\end{equation}
replaces the separate efficiency and resolution model used in
Ref.~\cite{Yuan:2025twx}.

\subsection{Background model and observed events}\label{sec:background}

The low-energy S2-only region is dominated by detector-induced
backgrounds rather than ordinary bulk electronic recoils. Following
the XENONnT analysis, the background model contains four components,
\begin{equation}
B_{ri} = B_{ri}^{\rm cath} + B_{ri}^{\rm DE} + B_{ri}^{\rm AE} + B_{ri}^{^{8}{\rm B}},
\label{eq:bkg_sum}
\end{equation}
where $B^{\rm cath}$ denotes cathode events,
$B^{\rm DE}$ delayed electrons,
$B^{\rm AE}$ accidental-electron pileup, and
$B^{^{8}{\rm B}}$ coherent elastic neutrino--nucleus scattering
(CE$\nu$NS) induced by solar $^{8}$B neutrinos.

The cathode component originates primarily from radioactive decays of
$^{210}$Pb and its daughters near the cathode electrode. It constitutes
the dominant residual background after the full event selection.
Delayed-electron events arise from electrons emitted following
preceding large S2 signals, whereas accidental-electron events are
formed by the pileup of otherwise unrelated single-electron signals.
The relative importance of the latter two components varies between
science runs because of changes in detector conditions and event
reconstruction. In particular, the higher S2 threshold in SR2
strongly suppresses the accidental-electron contribution.

We do not attempt to reconstruct these instrumental backgrounds from
first principles. Instead, we use the run- and bin-dependent
background templates released by the XENON Collaboration. This choice
retains the spectral information obtained from their sideband,
calibration, and detector-simulation studies while allowing us to
perform an independent reinterpretation for the solar-chameleon
signal.

\begin{figure}[t]
\centering
\includegraphics[width=0.99\linewidth]{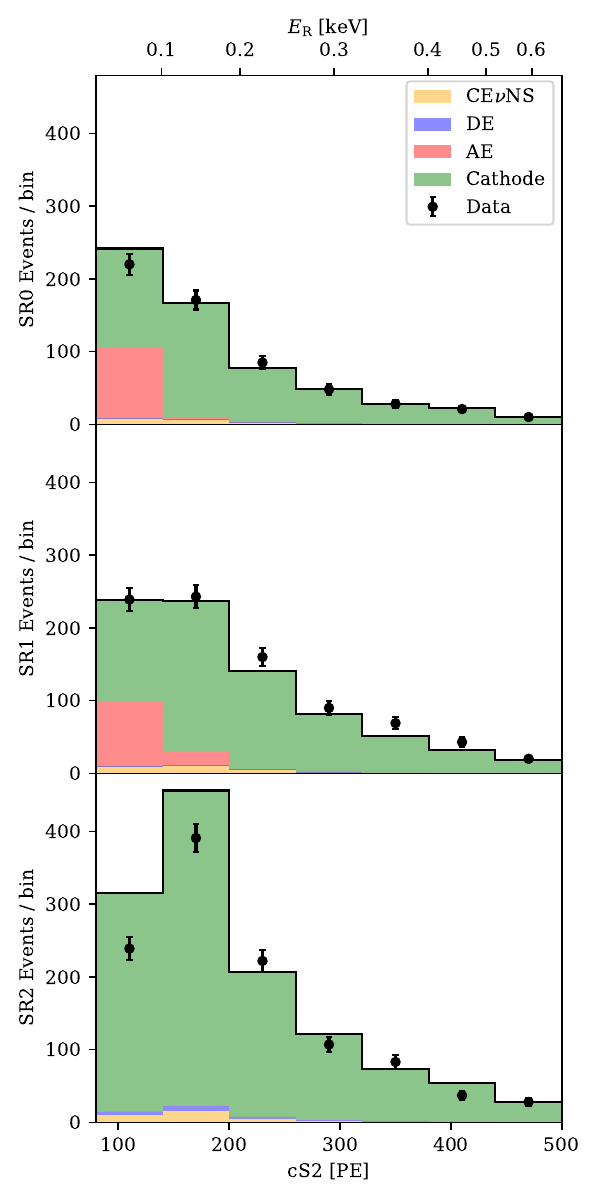}
\caption{
Observed XENONnT S2-only events and the expected background
contributions in the seven $\csTwo$ bins for SR0, SR1, and SR2.
The stacked histograms show the cathode, delayed-electron (DE),
accidental-electron (AE), and solar $^{8}$B CE$\nu$NS components,
while the black points denote the observed counts. The upper axis
indicates the approximate electronic-recoil energy corresponding to
the detector response. No significant upward excess is apparent in
any of the three science runs.
}
\label{fig:events_back}
\end{figure}

Fig.~\ref{fig:events_back} compares the released background model
with the observed events. The lower horizontal axis gives $\csTwo$ in
PE, while the upper axis gives an approximate electronic-recoil energy
for orientation. The seven $\csTwo$-bin centers are 110, 170, 230,
290, 350, 410, and 470~PE; their representative recoil energies are approximately 
0.11, 0.15, 0.22, 0.29, 0.36, 0.46, and 0.59~keV$_{ee}$,
respectively. These energies are not used as deterministic bin
assignments in the likelihood: the prediction is obtained from the
full response convolution in Eq.~\eqref{eq:response_folding}.
The total observed event numbers in the three science runs are
$N_{\rm obs}^{\rm SR0}=583, N_{\rm obs}^{\rm SR1}=864$, and $N_{\rm obs}^{\rm SR2}=1107$.
The corresponding total background expectations reported by XENONnT
are approximately
\begin{equation}
\begin{aligned}
B_{\rm SR0}&=600\pm80, \\
B_{\rm SR1}&=800\pm80, \\
B_{\rm SR2}&=1260\pm100 .
\end{aligned}
\end{equation}
before profiling the associated nuisance parameters. The first two
science runs are consistent with the background-only hypothesis,
while SR2 contains fewer events than the nominal expectation. The
latter is therefore a downward fluctuation rather than an excess that
could mimic a positive solar-chameleon contribution.
Instrumental backgrounds increase near threshold, where the accepted
solar-chameleon flux is also largest. Both effects enter the binned
likelihood below.

\subsection{Likelihood and Bayesian inference}
\label{sec:likelihood}
We follow the public XENONnT binning and background model, but construct our own likelihood for the solar-chameleon signal.
We construct a binned Poisson likelihood using the seven $\csTwo$ bins of each science run. For a given chameleon parameter point
$\boldsymbol{\Theta}$ and nuisance-parameter vector
$\boldsymbol{\eta}$, the expected number of events in run $r$ and
bin $i$ is
\begin{equation}
\mu_{ri}(\boldsymbol{\Theta},\boldsymbol{\eta})
=S_{ri}(\boldsymbol{\Theta},\boldsymbol{\eta})
+B_{ri}(\boldsymbol{\eta}).
\label{eq:expected_counts}
\end{equation}
The likelihood for the science data is
\begin{equation}
\mathcal{L}_{\rm data}
(\boldsymbol{\Theta},\boldsymbol{\eta})=\prod_{r} \prod_{i=1}^{7}
\frac{ \mu_{ri}^{\,N_{ri}^{\rm obs}} e^{-\mu_{ri}}}{ N_{ri}^{\rm obs}!},
\label{eq:poisson_likelihood}
\end{equation}
where the product over $r$ runs over SR0, SR1, and SR2.

Systematic uncertainties are incorporated through nuisance parameters.
For quantities with externally determined central values and
uncertainties, we adopt Gaussian constraint terms,
\begin{equation}
\pi(\eta_a) \propto \exp
\left[ -\frac{(\eta_a-\eta_{a,0})^2}{2\sigma_{\eta_a}^2} \right].
\label{eq:nuisance_prior}
\end{equation}
These account for the uncertainties associated with the background
normalizations and detector response. Where the released data provide
bin-dependent shape uncertainties, these are propagated directly into
the corresponding $\csTwo$ templates. The full likelihood is then
\begin{equation}
\mathcal{L} (\boldsymbol{\Theta},\boldsymbol{\eta}) =\mathcal{L}_{\rm data}
(\boldsymbol{\Theta},\boldsymbol{\eta})\prod_a \pi(\eta_a).
\label{eq:full_likelihood}
\end{equation}

For the five chameleon parameters, we perform a Bayesian analysis with
posterior density
\begin{equation}
p(\boldsymbol{\Theta},\boldsymbol{\eta}|\mathcal D)
=\frac{ \mathcal{L} (\mathcal D|\boldsymbol{\Theta},\boldsymbol{\eta})\,
\pi(\boldsymbol{\Theta})\, \pi(\boldsymbol{\eta}) }{\mathcal Z },
\label{eq:posterior}
\end{equation}
where $\pi(\boldsymbol{\Theta})$ denotes the model-parameter priors and
$\mathcal Z$ is the Bayesian evidence. Since the relevant couplings and
mass scales span many orders of magnitude, logarithmic parameters are
used for $\beta_e$, $\beta_\gamma$, $M_e$, and $\Lambda$, while $n$
is sampled directly. Explicitly, the sampled parameter vector is
\begin{equation}
\boldsymbol{\vartheta}=\left\{\log_{10}\beta_e,\, \log_{10}\beta_\gamma,\,
\log_{10}(M_e/{\rm eV}),\, \log_{10}(\Lambda/{\rm eV}),\, n \right\}.
\label{eq:sampling_parameters}
\end{equation}

\begin{table}[tb]
\caption{Priors, marginal posterior medians with 68\% credible intervals, and MAP values for the chameleon parameters.}
\label{tab:priors}
\begin{ruledtabular}
\begin{tabular}{lccc}
Parameter & Prior & Median (68\% C.I.) & MAP\\
\hline
$\log_{10}\beta_e$ & $\mathcal{U}(0,2)$ & $0.99\pm0.58$  & 1.22\\
$\log_{10}\beta_\gamma$ & $\mathcal{U}(0,15)$  & $5.8^{+2.7}_{-5.3}$ & 11.57\\
$\log_{10}(M_e/\mathrm{eV})$ & $\mathcal{U}(0,6)$  & $4.2^{+1.6}_{-0.82}$ & 0.55\\
$\log_{10}(\Lambda/\mathrm{eV})$ & $\mathcal{U}(-8,0)$  & $-4.1\pm2.3$ & -6.98\\
$n$ & $\mathcal{U}(0,6)$  & $3.0\pm1.7$ & 0.30\\
\end{tabular}
\end{ruledtabular}
\end{table}

The broad priors in Table~\ref{tab:priors} are chosen to expose the
parameter combinations identified by the XENONnT data alone. They are
not intended to encode external laboratory or astrophysical
constraints. This distinction matters for $\beta_e$, $\Lambda$, and
$n$, whose marginal posteriors remain weakly identified and therefore
retain substantial prior dependence. By contrast, the data constrain
a narrow direction in the $\beta_\gamma$--$M_e$ plane corresponding
approximately to constant $\beff$. We consequently regard $\beff$ as
the principal derived parameter of the S2-only analysis.

We explore the posterior using an affine-invariant Markov-chain Monte
Carlo sampler through the affine-invariant ensemble sampler \texttt{emcee}~\cite{Foreman-Mackey:2012any}, and analyzing the chains using the \texttt{GetDist} package~\cite{Lewis:2019xzd}. Convergence is assessed from the integrated
autocorrelation time of each sampled parameter, and the initial
portion of each chain is discarded as burn-in. Unless stated
otherwise, all marginalized intervals reported below are obtained
from the converged posterior samples.

The likelihood is evaluated jointly for all 21 run--bin combinations.
Run-dependent response and background nuisance parameters are retained
throughout the sampling and integrated out when constructing the
marginal posteriors. This procedure preserves the statistical
independence of SR0, SR1, and SR2 while propagating their different
thresholds and detector conditions into the final constraint.

Two complementary summaries of the inference are used in
Sec.~\ref{sec:results}. First, we identify the maximum-a-posteriori
(MAP) parameter point,
\begin{equation}
\boldsymbol{\Theta}_{\rm MAP} = \arg\max_{\boldsymbol{\Theta}}
p(\boldsymbol{\Theta}|\mathcal D),
\end{equation}
which is used only to illustrate the characteristic signal shape in
$\csTwo$ space and should not be interpreted as evidence for a
chameleon signal. Second, we marginalize over the remaining model and
nuisance parameters to derive constraints on the effective disformal
coupling
\begin{equation}
\beff = \beta_\gamma 
\left( \frac{{\rm eV}}{M_e} \right)^4
\end{equation}
as a function of the conformal electron coupling $\beta_e$.
The quoted upper bounds correspond to the 95\% posterior credible level unless explicitly stated otherwise.

\section{Results}
\label{sec:results}

We now present the constraints on the generalized chameleon model
obtained from the XENONnT ionization-only data. As described in
Sec.~\ref{sec:xenon_analysis}, the theoretical solar-chameleon
electronic-recoil spectrum is folded with the run-dependent detector
response and compared with the observed events in the seven $\csTwo$
bins of SR0, SR1, and SR2. We first illustrate the signal morphology
at the maximum-a-posteriori (MAP) point, then discuss the marginalized
posterior distributions of the five chameleon parameters, and finally
present the resulting constraint in the $\beff$--$\beta_e$ plane.

\subsection{Maximum-a-posteriori signal}
\label{sec:map_result}

The posterior is maximized at
\begin{equation}
\begin{split}
\log_{10}\beta_e &= 1.22,\\
\log_{10}\beta_\gamma &= 11.57,\\
\log_{10}\left(\frac{M_e}{\mathrm{eV}}\right) &= 0.55,\\
\log_{10}\left(\frac{\Lambda}{\mathrm{eV}}\right) &= -6.98,\\
n &= 0.30 .
\end{split}
\label{eq:map_point}
\end{equation}
Equivalently, the dimensional parameters at this point are
approximately $\beta_e \simeq 1.7\times10^{1}, \beta_\gamma \simeq 3.7\times10^{11}, M_e \simeq 3.5~\mathrm{eV}$, and $\Lambda \simeq 1.0\times10^{-7}~\mathrm{eV}$.
The MAP point provides a representative signal morphology within the
multidimensional posterior. The ionization-only data contain no
significant upward excess over the background expectation, so this
point is not evidence for a nonzero chameleon contribution.

\begin{figure}[t]
\centering
\includegraphics[width=0.98\linewidth]{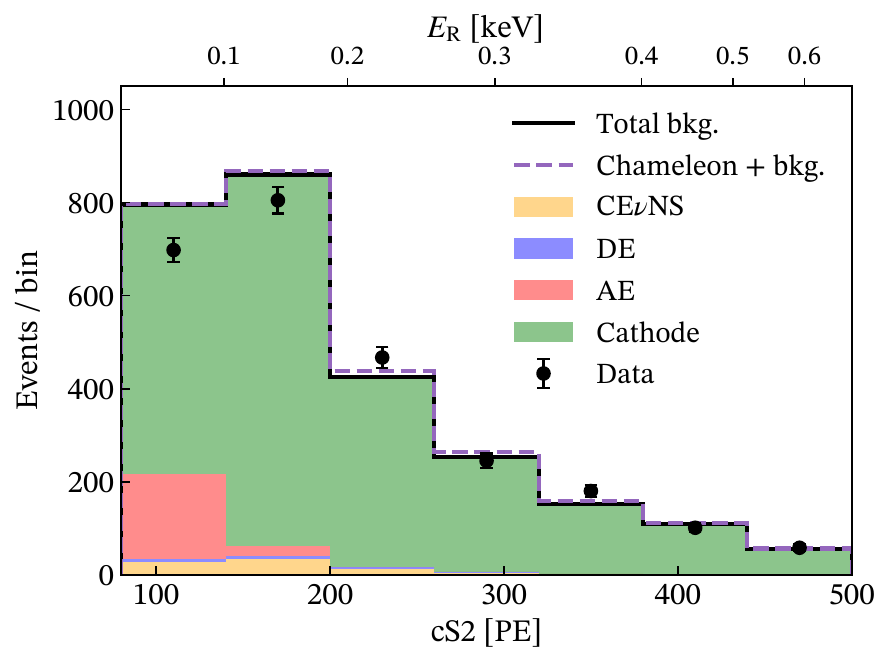}
\caption{
Observed XENONnT ionization-only events compared with the background
model and the solar-chameleon prediction evaluated at the MAP point
of Eq.~\eqref{eq:map_point}. The stacked histograms show the expected background
contributions and the response-folded chameleon signal. The total
prediction is shown by the solid curve, and the black points are the
observed counts in the seven $\csTwo$ bins. The three runs are fitted
separately in the likelihood.
}
\label{fig:fitting}
\end{figure}

Fig.~\ref{fig:fitting} compares the response-folded MAP signal plus
background with the data. The chameleon contribution is concentrated
toward the lowest $\csTwo$ bins, reflecting both the rapidly varying
low-energy solar-chameleon spectrum and the detector response in the
few-electron regime.

The three science runs respond differently to the same underlying
solar-chameleon spectrum because of their run-dependent response
matrices, thresholds, exposures, and background compositions.
In particular, SR2 has the largest exposure but also a higher
low-$\csTwo$ threshold and a different instrumental-background
environment. Treating the three science runs independently in the
likelihood therefore preserves information that would be lost by
combining them into a single exposure-weighted spectrum.

\subsection{Posterior constraints on the chameleon parameters}
\label{sec:posterior_results}

\begin{figure*}[t]
\centering
\includegraphics[width=0.90\linewidth]{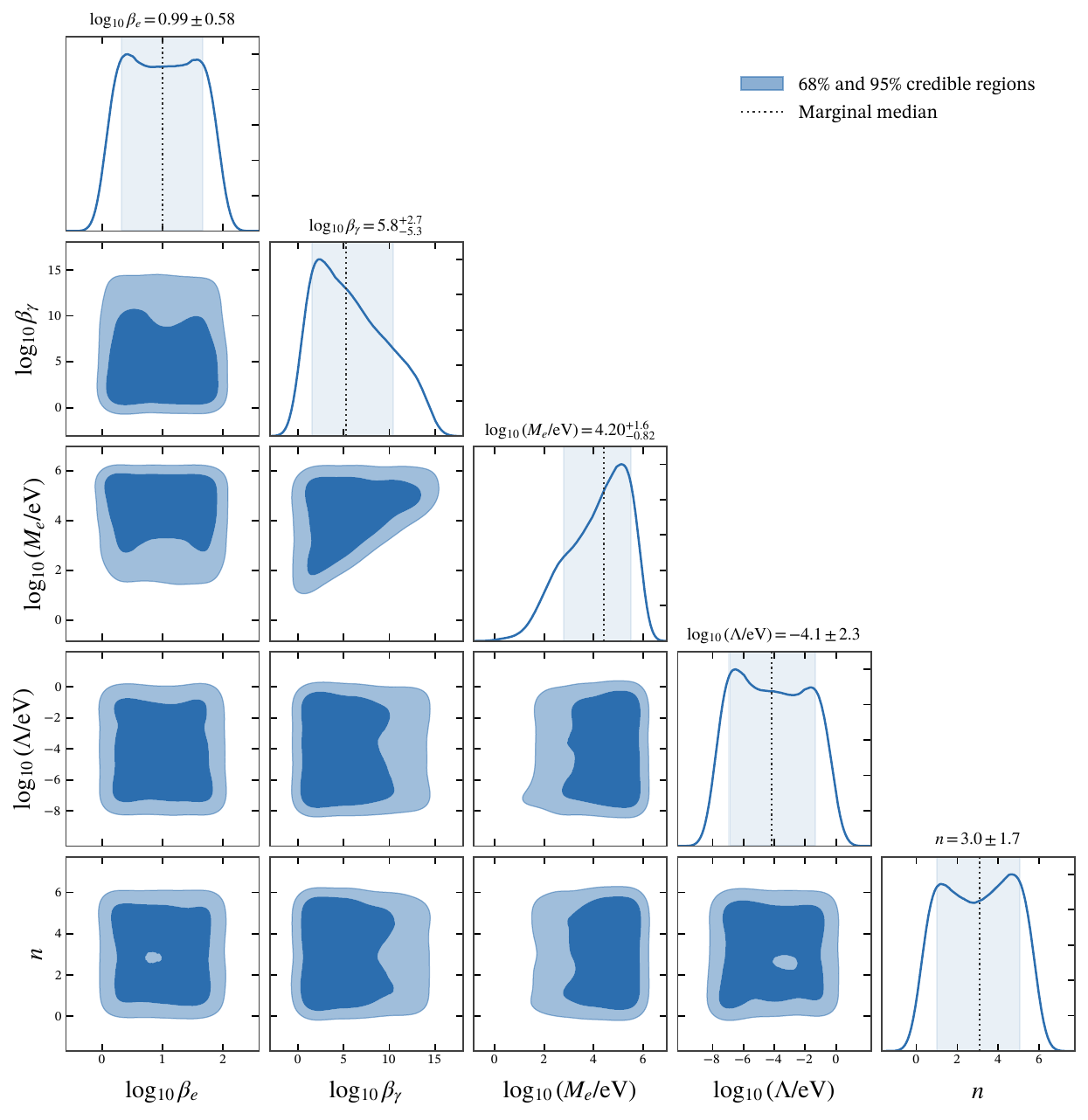}
\caption{
Marginalized posterior distributions of the five chameleon
parameters
$\{\beta_e,\beta_\gamma,M_e,\Lambda,n\}$
obtained from the XENONnT ionization-only data.
The diagonal panels show the one-dimensional marginalized
distributions, while the off-diagonal panels show the corresponding
two-dimensional distributions. Dotted lines mark the
marginal medians, the dark and light contours enclose 68\%
and 95\% posterior probability, respectively.
}
\label{fig:corner}
\end{figure*}

The marginalized posterior distributions of the five model
parameters are shown in Fig.~\ref{fig:corner}. The diagonal panels
display the one-dimensional marginalized posteriors, while the
off-diagonal panels show the corresponding two-dimensional credible
regions. The MAP point of Eq.~\eqref{eq:map_point} is indicated for
reference in Table~\ref{tab:priors}.

The posterior structure reflects the different roles played by the
five parameters in solar production and terrestrial detection.
The photon coupling $\beta_\gamma$ controls the normalization
of the solar chameleon flux, whereas the electron disformal scale
$M_e$ determines the strength of the disformal detection channel.
As discussed in Sec.~\ref{sec:recoil_spectrum}, the corresponding
event rate approximately scales as
\begin{equation}
R_{\rm dis} \propto \beta_\gamma^2 M_e^{-8}.
\end{equation}
Consequently, the data are expected to constrain most directly the combination $\beff \equiv \beta_\gamma \left(\frac{\mathrm{eV}}{M_e}\right)^4$,
rather than $\beta_\gamma$ and $M_e$ independently. The resulting
degeneracy appears as an elongated posterior structure in
the $\beta_\gamma$--$M_e$ plane.

The conformal electron coupling $\beta_e$ enters the detection rate
through the photoabsorption-like contribution $\sigma_{\phi e}^{\rm conf} \propto \beta_e^2 $,
and therefore controls a physically distinct contribution to the
signal. The relative importance of the conformal and disformal terms
changes across the posterior, which motivates presenting the final
constraint in the $\beff$--$\beta_e$ plane.

The parameters $\Lambda$ and $n$ affect the chameleon effective mass
and hence the propagation and production conditions inside the Sun.
Over the region to which the present data are sensitive, their
one-dimensional posteriors, together with that of $\beta_e$, remain
broad and largely prior dominated. The numerical summaries in
Table~\ref{tab:priors} should therefore not be read as independent
measurements of these three parameters. The pronounced
$\beta_\gamma$--$M_e$ ridge is the statistically better determined
feature of the five-dimensional posterior and is captured by
$\beff$.

The MAP point need not coincide with the peaks of the one-dimensional
marginal distributions. Extended degeneracies can contain more
posterior volume than the neighborhood of the global maximum. We use
the MAP point in Fig.~\ref{fig:fitting} only as a diagnostic of the
spectral fit; no model-selection statistic or discovery significance
is assigned to it.

\subsection{Constraint on the effective coupling}
\label{sec:effective_constraint}

The approximate degeneracy between $\beta_\gamma$ and $M_e$ is most
conveniently expressed in terms of the effective coupling $\beff$.
We therefore marginalize over the remaining model parameters and
nuisance parameters and derive the 95\% Bayesian upper limit on
$\beff$ as a function of the conformal electron coupling $\beta_e$.

\begin{figure}[t]
\centering
\includegraphics[width=0.98\linewidth]{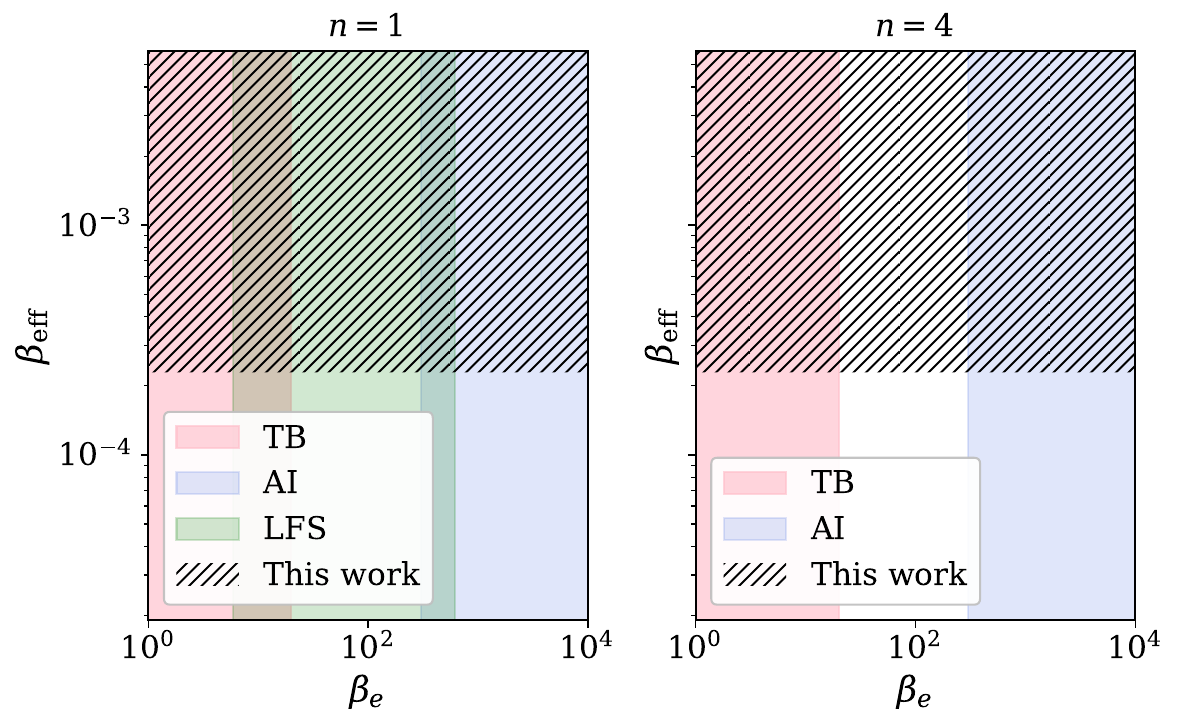}
\caption{
Constraints from the XENONnT ionization-only data in the
$\beff$--$\beta_e$ plane for $n=1$ (left) and $n=4$ (right), after
marginalization over the remaining model and nuisance parameters.
The hatched region is excluded at the 95\% Bayesian credible level (C.L.).
In the disformal-dominated regime, the limit is nearly independent of
$\beta_e$ and approaches $\log_{10}\beff<-3.64$.
Shaded regions show constraints from atom interferometry (AI),
levitated force sensors (LFS), and torsion-balance experiments (TB).
The LFS bound does not apply for $n=4$.
}
\label{fig:betaeff_limit}
\end{figure}

Figure~\ref{fig:betaeff_limit} shows the resulting exclusion regions.
For small $\beta_e$, the conformal contribution to the
chameleon--electron cross section is negligible, and the predicted
event rate is dominated by the disformal interaction. Since this
contribution scales approximately as $\beff^2$, the corresponding
upper bound on $\beff$ is nearly independent of $\beta_e$ in this
regime.
For comparison, Fig.~\ref{fig:betaeff_limit} also shows existing
constraints from atom interferometry, levitated force sensors, and
torsion-balance experiments
~\cite{Wagner:2012ui,Jenke:2014yel,Burrage:2017qrf,Yin:2022geb,
MICROSCOPE:2022doy}.

As $\beta_e$ increases, the conformal channel contributes an
increasing fraction of the predicted event rate. The allowed
disformal contribution must then decrease in order to remain
compatible with the observed counts, producing the downward turn of
the exclusion boundary at large $\beta_e$.

For both benchmark values of $n$, the ionization-only data give an
asymptotic bound $\log_{10}\beff < -3.64$.
This constraint is substantially weaker than the bound obtained in
Ref.~\cite{Yuan:2025twx}, whose $\log_{10}\beff < -6.91 $.
The difference reflects the distinct experimental regimes probed by
the two analyses. The previous result was derived from a reconstructed
electronic-recoil spectrum at higher energies, whereas the present
analysis uses the few-electron S2-only sample. Although the latter
extends the recoil threshold to lower energies, the gain in kinematic
reach is offset by the large instrumental background and the low
signal acceptance close to threshold. For the solar-chameleon spectrum
considered here, these effects dominate over the benefit of the lower
energy threshold and lead to a weaker constraint on $\beff$.

The present analysis nevertheless probes the signal in a qualitatively
different observable regime. Rather than assigning each deposited
energy to a single reconstructed energy, the true ER spectrum is folded
through the run-dependent detector response before the likelihood is
evaluated. This treatment is required near the ionization threshold,
where fluctuations in the number of extracted electrons produce a
broad mapping between $E_R$ and $\csTwo$. The resulting bound therefore
provides a direct constraint from the XENONnT ionization-only data,
with the detector response and the different SR0, SR1, and SR2
conditions treated explicitly.

The constraints in Fig.~\ref{fig:betaeff_limit} complement those from
the conventional ER analysis by testing the low-energy, few-electron
part of the solar-chameleon signal. Their physical implications and
the role of the low-energy detector response are discussed further in
Sec.~\ref{sec:discussion}.

\section{Discussion and Conclusions}
\label{sec:discussion}

The XENONnT ionization-only data extend the electronic-recoil coverage
down to approximately $0.04~{\rm keV}_{ee}$, where the detector
response must be included explicitly in the signal prediction. In this
few-electron regime, fluctuations in ionization production, electron
extraction, and proportional scintillation lead to a broad relation
between the deposited energy $E_R$ and the measured $\csTwo$ signal.
We therefore fold the recoil spectrum through the separate SR0, SR1,
and SR2 response matrices before evaluating the likelihood.

The lower recoil threshold does not necessarily yield a stronger
constraint on the chameleon couplings. Although the S2-only analysis
is sensitive to a larger fraction of the low-energy solar-chameleon
flux, the disformal absorption cross section scales as $E_R^4$.
Its contribution consequently falls rapidly toward lower recoil
energies. The lowest-$\csTwo$ bins also contain the largest
instrumental backgrounds, dominated by cathode events together with
delayed and accidental electrons. The combination of the suppressed
low-energy interaction rate and the increased background explains why
the present bound is weaker than that obtained from the reconstructed
ER spectrum in Ref.~\cite{Yuan:2025twx}, despite the larger exposure
and lower threshold.
Keeping the three science runs separate is important for this
comparison. SR2 provides the largest exposure, but its S2 threshold is
higher and its instrumental-background composition differs from those
of SR0 and SR1. The observed SR2 counts also lie below the nominal
background expectation. A positive chameleon contribution cannot
account for this downward fluctuation, while the joint likelihood
simultaneously constrains any signal contribution allowed by the
low-$\csTwo$ bins of SR0 and SR1. Combining the three runs into a
single exposure-weighted spectrum would discard this run-dependent
information.

The posterior constrains one combination of couplings more directly
than the individual Lagrangian parameters. Primakoff production scales
approximately as $\beta_\gamma^2$, whereas disformal absorption scales
as $M_e^{-8}$. The corresponding contribution to the event rate is
therefore proportional to $\beta_\gamma^2 M_e^{-8}=\beff^2$.
The elongated posterior in the $\beta_\gamma$--$M_e$ plane follows
approximately a line of constant $\beff$. By comparison, the
marginalized distributions of $\beta_e$, $\Lambda$, and $n$ remain
broad over the adopted prior ranges. The MAP point is useful for
illustrating the spectral form of an allowed signal, but it should not
be interpreted as a measurement of all five parameters or as evidence
for a solar-chameleon contribution.

Constraints on either $\beta_\gamma$ or $M_e$ from other experiments
can, in principle, reduce this degeneracy, although their applicability
is model dependent. Bounds on $\beta_\gamma$ inferred from solar
chameleon searches may depend on the assumed production region and
solar magnetic-field profile. Collider limits on $M_e$, on the other
hand, are commonly obtained for effectively massless scalars in the
absence of environmental screening. Translating either class of bounds
to the generalized chameleon scenario considered here requires the
same propagation, screening, and effective-field-theory assumptions
to remain valid.

The production model also defines the scope of our result. We include
transverse Primakoff production in the screened electric fields of the
solar plasma and photon--chameleon conversion in macroscopic magnetic
fields. Primakoff production dominates most of the energy interval
relevant to the present fit, reducing the dependence of the predicted
flux on the less certain solar magnetic-field profile. Longitudinal
plasma modes are not included. If such modes provide an additional
positive low-energy flux without modifying the detection process, the
predicted event rate would increase and the inferred upper bound would
become stronger. A consistent treatment of this contribution requires
a dedicated calculation for the generalized chameleon model.

Further progress in the ionization-only channel will depend strongly
on control of the few-electron response. More precise measurements of
the low-energy ionization yield, electron extraction, and
single-electron gain would reduce the detector-response uncertainty
close to threshold. Lower and more stable S2 thresholds, together with
improved rejection of cathode and delayed-electron events, would also
increase the useful sensitivity of the first $\csTwo$ bins. Data from
other target materials could provide complementary information because
the conformal contribution follows the target photoabsorption response,
whereas the disformal contribution has a different energy and atomic
dependence. Comparisons among xenon, argon, germanium, and silicon
targets may therefore help separate the two interaction channels if a
low-energy excess were observed.

We have analyzed the $7.83~{\rm tonne\,yr}$ XENONnT ionization-only
data set using a response-folded solar-chameleon signal and a joint
Poisson likelihood for the seven $\csTwo$ bins of SR0, SR1, and SR2.
In the disformal-dominated regime, we obtain
\begin{equation}
    \log_{10}\beff < -3.64
    \qquad
    (95\%~{\rm C.L.}) .
\end{equation}
For comparison, the higher-energy reconstructed-ER analysis gave
$\log_{10}\beff<-6.91$. The weaker S2-only bound follows from the
$E_R^4$ dependence of the disformal absorption cross section and the
large instrumental background near the ionization threshold. The
ionization-only data nevertheless test the solar-chameleon spectrum in
a distinct sub-keV regime and provide a detector-level framework for
future searches that exploit few-electron signals.

\begin{acknowledgments}
The author thanks Sunny Vagnozzi and Fei Gao for helpful discussions, and Shengyang Shi for sharing data and discussing the analysis. This work was supported by the University of Trento and the Provincia Autonoma di Trento through the UniTrento Internal Call for Research 2023 grant ``Searching for Dark Energy off the beaten track'' (DARKTRACK, grant agreement No.~E63C22000500003, PI: Sunny Vagnozzi), and by the Istituto Nazionale di Fisica Nucleare through the Commissione Scientifica Nazionale 4 initiative ``Quantum Fields in Gravity, Cosmology and Black Holes'' (FLAG).
\end{acknowledgments}

\bibliography{solarchameleon}

\end{document}